\documentclass[twocolumn]{aastex62}
\usepackage{natbib}
\usepackage{epsfig}
\usepackage{graphicx}
\usepackage{subfigure}
\usepackage{float}
\usepackage{amsmath}
\usepackage{color}
\usepackage{amssymb}
\usepackage{amsfonts}
\usepackage{apjfonts}
\newcommand{\PMO}{Purple Mountain Observatory, Chinese Academy of Sciences, Nanjing 210033, China}
\newcommand{\GXU}{Guangxi Key Laboratory for Relativistic Astrophysics, Guangxi University, Nanning 530004, China}
\newcommand{\UCAS}{University of Chinese Academy of Sciences, Beijing 100049, China}
\newcommand{\UA}{Department of Physics, The Applied Math Program, and Department of Astronomy, The University of Arizona, Tucson, AZ 85721, USA}

\shortauthors{Wei \& Melia}

\begin{document}

\title{Model-independent Distance Calibration and Curvature Measurement using Quasars and Cosmic Chronometers}

\correspondingauthor{Jun-Jie Wei, Fulvio Melia}
\email{jjwei@pmo.ac.cn, fmelia@email.arizona.edu}

\author{Jun-Jie Wei}
\affiliation{\PMO}
\affiliation{\GXU}
\affiliation{\UCAS}

\author{Fulvio Melia \thanks{John Woodruff Simpson Fellow.}}
\affiliation{\UA}

\begin{abstract}
We present a new model-independent method to determine the spatial curvature and to mitigate the circularity
problem affecting the use of quasars as distance indicators. The cosmic-chronometer measurements are used to
construct the curvature-dependent luminosity distance $D^{\rm cal}_{L}(\Omega_{K},z)$ using a polynomial fit.
Based on the reconstructed $D^{\rm cal}_{L}(\Omega_{K},z)$ and the known ultraviolet versus X-ray luminosity
correlation of quasars, we simultaneously place limits on the curvature parameter $\Omega_{K}$ and the parameters
characterizing the luminosity correlation function. This model-independent analysis suggests that a mildly closed
Universe ($\Omega_{K}=-0.918\pm0.429$) is preferred at the $2.1\sigma$ level. With the calibrated luminosity
correlation, we build a new data set consisting of 1598 quasar distance moduli, and use these calibrated
measurements to test and compare the standard $\Lambda$CDM model and the $R_{\rm h}=ct$ universe. Both models
account for the data very well, though the optimized flat $\Lambda$CDM model has one more free parameter than
$R_{\rm h}=ct$, and is penalized more heavily by the Bayes Information Criterion. We find that $R_{\rm h}=ct$
is slightly favoured over $\Lambda$CDM with a likelihood of $\sim57.7\%$ versus 42.3\%.

\end{abstract}


\section{Introduction}
\label{sec:Introduction}
Quasars are the most luminous persistent sources in the Universe, detected up to redshifts $z\sim7.5$
\citep{2011Natur.474..616M,2015Natur.518..512W,2018Natur.553..473B}. It is generally accepted that
the ultraviolet (UV) photons of active galactic nuclei are emitted by an accretion disk, while the X-rays
are Compton upscattered photons from a hot corona above the disk. A non-linear relation between the UV
and X-ray monochromatic luminosities of quasars has been known for over three decades \citep{1986ApJ...305...83A},
but only recently has the uncomfortably large dispersion in the correlation been mitigated by refining the
selection criteria and flux measurements \citep{2015ApJ...815...33R,2019NatAs...3..272R,2016ApJ...819..154L,2017A&A...602A..79L}.
This offers the possibility of using quasars as a complementary cosmic probe at high-$z$. Some cosmological
constraints have been obtained based on this refined correlation
\citep{2015ApJ...815...33R,2019NatAs...3..272R,2016IJMPD..2550060L,2019arXiv190901400K,2019A&A...628L...4L,2019MNRAS.489..517M}.
There is a circularity problem when treating high-$z$ quasars as relative standard candles, however. The
problem arises from the fact that, given the lack of very low-$z$ quasars, the correlation between the
X-ray and UV luminosities must be established assuming a background cosmology. All of the previous
applications of the luminosity correlation attempting to overcome the circularity problem have used
simultaneous multi-parameter fitting in the context of specifically selected models. One of the principal
limitations of this approach is that none of the chosen models may actually be the true cosmology.

These hurdles notwithstanding, there is significant motivation to use these data for cosmological
studies because high-$z$ quasars extend our reach well beyond other kinds of sources, such as Type Ia supernovae (SNe Ia),
which may be seen only as far as $z\sim 1.8$. Quasars may therefore help to shed light on one of the most
pressing issues in modern cosmology, i.e., the spatial curvature of the Universe. Estimating whether the
Universe is open, flat, or closed is crucial for us to understand the evolution of the Universe and the
nature of dark energy \citep{2006JCAP...12..005I,2007JCAP...08..011C,2007PhRvD..75d3520G,2008JCAP...12..008V}.
Any significant deviation from zero spatial curvature would have a profound impact on the inflationary
paradigm and its underlying physics \citep{2005ApJ...633..560E,2006PhRvD..74l3507T,2007ApJ...664..633W,2007PhLB..648....8Z}.

Current cosmological observations strongly favor a spatially flat Universe, e.g., the combined
Planck2018 cosmic microwave background (CMB) and baryon acoustic oscillation measurements, which
suggest that $\Omega_{K}=0.001\pm0.002$ \citep{2018arXiv180706209P}.\footnote{The combination of the
Planck2018 CMB temperature and polarization power spectra data slightly favor a mildly closed Universe,
i.e., $\Omega_{K}=-0.044^{+0.018}_{-0.015}$ \citep{2018arXiv180706209P}. Other recent works showed that
the Planck2015 CMB anisotropy data also favor a mildly closed Universe (see
\citealt{2018ApJ...864...80O,2018ApJ...866...68O} and references therein).} These constraints, however,
are based on the pre-assumption of a specific cosmological model (e.g., the standard $\Lambda$CDM model).
Because of the strong degeneracy between spatial curvature and the equation of state of dark energy,
it is rather difficult to constrain the two quantities simultaneously. In general, dark energy is assumed
to be a cosmological constant for the estimation of curvature, or conversely, the Universe is assumed
to be flat in a dark energy analysis. But a simple flatness assumption may result in an incorrect reconstruction
of the dark energy equation of state, even if the real curvature is very tiny \citep{2007JCAP...08..011C},
and a cosmological constant assumption may lead to confusion between $\Lambda$CDM and a
dynamical dark-energy model \citep{2008JCAP...12..008V}. Therefore, it would be better to measure
the curvature parameter by purely geometrical and model-independent methods. A non-exhaustive list
of previous works attempting to measure spatial curvature in a model-independent way includes
\cite{2006ApJ...637..598B,2007JCAP...08..011C,2010PhRvD..81h3537S,2011arXiv1102.4485M,2014ApJ...789L..15L,2014PhRvD..90b3012S,
2015PhRvL.115j1301R,2016PhRvD..93d3517C,2016ApJ...833..240L,2018ApJ...854..146L,2016ApJ...828...85Y,2017JCAP...01..015L,
2017ApJ...839...70L,2017JCAP...03..028R,2017ApJ...847...45W,2017ApJ...838..160W,2017ApJ...834...75X,2018JCAP...03..041D,
2018ApJ...868...29W,2018ApJ...856....3Y,2019PDU....24..274C,2019ApJ...881..137R,2019MNRAS.483.1104Q}.

In this paper, we propose a new model-independent method to simultaneously calibrate the UV versus X-ray luminosity
correlation for quasars and to determine the curvature of the Universe. Using a polynomial fitting technique
\citep{2019MNRAS.486L..46A}, one can reconstruct a continuous $H(z)$ function representing the expansion rate
measurements using cosmic chronometers without the pre-assumption of any particular cosmological model. The
co-moving distance function can then be directly derived by integrating the reconstructed $H(z)$ function. Then,
with the curvature parameter $\Omega_{K}$ taken into consideration, the co-moving distance can be transformed into
the curvature-dependent luminosity distance $D^{\rm cal}_{L}(\Omega_{K},z)$. Finally, by combining
$D^{\rm cal}_{L}(\Omega_{K},z)$ with the redshift and flux measurements of quasars, we obtain model-independent
constraints on the spatial curvature and the parameters characterizing the quasar luminosity relation. Refining
their selection technique to avoid the inclusion of possible contaminants, \cite{2019NatAs...3..272R} obtained
a final catalogue of 1598 quasars with reliable measurements of both the intrinsic X-ray and UV emissions.
We adopt this high-quality quasar sample covering the redshift range $0.04<z<5.1$ for the analysis reported in
this paper.

The rest of the paper is organized as follows. In Section~\ref{sec:method}, we introduce the model-independent
method used to calibrate the quasar luminosity relation and determine the curvature parameter. The calibrated
luminosity relation is then used to test different cosmological models. The outcome of our model comparison
will be described in Section~\ref{sec:constraint}, followed by our conclusions in Section~\ref{sec:summary}.

\section{Methodology}
\label{sec:method}
\subsection{Curvature-dependent Distance from Cosmic-chronometer Measurements}
\begin{figure}
\vskip-0.3in
\centerline{\includegraphics[keepaspectratio,clip,width=0.5\textwidth]{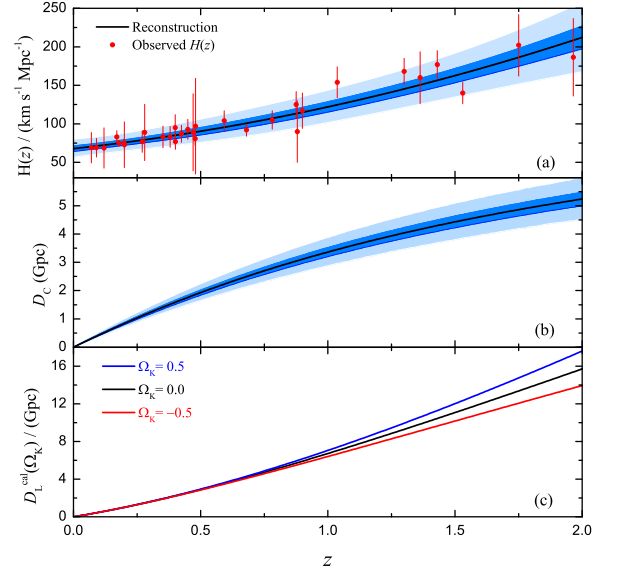}}
\vskip-0.3in
\caption{Panel (a): measured expansion rates from cosmic-chronometer measurements (solid points with the
vertical error bars) and the reconstructed Hubble parameter function $H(z)$ (solid line) from the data.
Panel (b): the reconstructed co-moving distance function $D_{C}(z)$ (solid line). The shaded areas in both
Panels (a) and (b) represent the $1\sigma$ and $3\sigma$ confidence regions of the corresponding
reconstructions. Panel (c): indicative dependence of the luminosity distance $D^{\rm cal}_{L}$ derived
from the $D_{C}(z)$ function on the spatial curvature parameter $\Omega_{K}$.}
\label{f1}
\end{figure}

The expansion rate of the Universe, $H(z)\equiv\dot{a}/a$, where $a(t)=1/(1+z)$, can be obtained directly
from the redshift-time derivative $dz/dt$ using $H(z)=-\frac{1}{1+z}\frac{dz}{dt}$ at any redshift $z\not=0$.
For this purpose, the differential age evolution of passively evolving galaxies can be used to measure the
expansion rate $H(z)$ in a cosmology-independent way \citep{2002ApJ...573...37J}. These galaxies are commonly
referred to as `cosmic chronometers.' The most recent sample of 31 cosmic-chronometer measurements
(see \citealt{2018ApJ...868...29W} and references therein) is shown in Figure~\ref{f1}(a). To avoid the
circularity problem, \citet{2019MNRAS.486L..46A} proposed a model-independent technique to reconstruct
a reasonable $H(z)$ function that best approximates the discrete cosmic-chronometer data. Following
these authors, we fit the $H(z)$ measurements employing a B$\rm \acute{e}$zier parametric curve of degree $n$:
\begin{eqnarray}
H_{n}(z)&=&\sum_{d=0}^{n}\beta_{d}h_{n}^{d}(z)\nonumber \\
h_{n}^{d}(z)&\equiv&\frac{n!(z/z_{\rm m})^{d}}{d!(n-d)!}\left(1-\frac{z}{z_{\rm m}}\right)^{n-d}\;,
\end{eqnarray}
where $z_{\rm m}$ is the maximum redshift of the cosmic-chronometer data set and the $\beta_{d}$ are positive
coefficients of the linear combination of Bernstein basis polynomials $h_{n}^{d}(z)$ in the range $0\leq z\leq z_{\rm m}$.
For $z=0$ and $d=0$, it is easy to identify $\beta_{0}\equiv H_{0}$. Since the high value of $n$ ($n>2$) would
lead to an oscillatory behavior of the approximating function, we adopt $n=2$ in fitting the discrete $H(z)$ data \citep{2019MNRAS.486L..46A}.
The reconstructed $H(z)$ function (solid line) with $1\sigma$ and $3\sigma$ confidence regions (shaded areas)
are plotted in Figure~\ref{f1}(a). The best-fitting parameters are $H_{0}=67.76\pm3.68$, $\beta_{1}=103.33\pm11.16$,
and $\beta_{2}=208.45\pm14.29$ (all in units of km $\rm s^{-1}$ $\rm Mpc^{-1}$). The Hubble constant
$H_{0}$ obtained here is in good agreement with the value inferred from {\it Planck} ($H_{0}=67.4\pm0.5$ km $\rm s^{-1}$
$\rm Mpc^{-1}$; \citealt{2018arXiv180706209P}), and is also compatible at the $1.6\sigma$ level with the estimate
based on local distance measurements ($H_{0}=74.03\pm1.42$ km $\rm s^{-1}$ $\rm Mpc^{-1}$; \citealt{2019ApJ...876...85R}).
For consistency, we use this best-fit value of $H_{0}$ for the distance estimations in the following analysis.

The line-of-sight co-moving distance
\begin{equation}
D_{C}(z)=c\int^{z}_{0}\frac{dz'}{H_{2}(z')}
\label{eq:dp}
\end{equation}
\citep{1999astro.ph..5116H} can then be derived by integrating the $H_{2}(z)$ function with respect to redshift
(solid line in Figure~\ref{f1}b). Since the error propagation is complicated, we estimate uncertainties in
$D_{C}(z)$ based on 10,000 Monte Carlo simulations utilizing the uncertainties in the coefficients $H_{0}$,
$\beta_{1}$, and $\beta_{2}$. The shaded areas in Figure~\ref{f1}(b) represent the $1\sigma$ and $3\sigma$
uncertainties taking into account the spread of all the Monte Carlo simulation results.

With the reconstructed co-moving distance function $D_{C}(z)$, as well as its $1\sigma$ uncertainty $\sigma_{D_{C}}$,
the curvature-dependent luminosity distance $D^{\rm cal}_{L}$ can then be expressed as
\begin{equation}\label{DL_H}
\frac{D^{\rm cal}_{L}(\Omega_{K},z)}{(1+z)} = \left\lbrace \begin{array}{lll} \frac{D_{H}}{\sqrt{\Omega_{K}}}\sinh\left[\sqrt{\Omega_{K}}\frac{D_{C}(z)}{D_{H}}\right]~~~~~{\rm for}~~\Omega_{K}>0\\
                                         D_{C}(z)~~~~~~~~~~~~~~~~~~~~~~~~~~~~~~~{\rm for}~~\Omega_{K}=0 \\
                                         \frac{D_{H}}{\sqrt{|\Omega_{K}|}}\sin\left[\sqrt{|\Omega_{K}|}\frac{D_{C}(z)}{D_{H}}\right]~\,{\rm for}~~\Omega_{K}<0\;,\\
\end{array} \right.
\end{equation}
with its corresponding uncertainty
\begin{equation}
\sigma_{D^{\rm cal}_{L}} = \left\lbrace \begin{array}{lll} (1+z)\cosh\left[\sqrt{\Omega_{K}}\frac{D_{C}(z)}{D_{H}}\right]\sigma_{D_{C}}~~~\;{\rm for}~~\Omega_{K}>0\\
                                         (1+z)\sigma_{D_{C}}~~~~~~~~~~~~~~~~~~~~~~~~~~~~~~~~{\rm for}~\Omega_{K}=0 \\
                                         (1+z)\cos\left[\sqrt{|\Omega_{K}|}\frac{D_{C}(z)}{D_{H}}\right]\sigma_{D_{C}}~~\;{\rm for}~~\Omega_{K}<0\;,\\
\end{array} \right.
\end{equation}
where $D_{H}=c/H_0$ is the Hubble distance. In Figure~\ref{f1}(c), we illustrate the dependence of $D^{\rm cal}_{L}$
(derived from cosmic-chronometer measurements) on the spatial curvature $\Omega_{K}$.

\subsection{Distance Calibration and Curvature Measurement}
We are now in position to use $D^{\rm cal}_{L}(\Omega_{K},z)$ derived from cosmic-chronometer measurements
to calibrate the non-linear relation between the UV and X-ray luminosities of quasars, which is commonly
written using the following ansatz:
\begin{equation}
\log_{10}L^{\rm cal}_{\rm X}=\gamma \log_{10}L^{\rm cal}_{\rm UV}+\kappa\;.
\label{eq:L-relation}
\end{equation}
We re-write this to bring out its explicit dependence on the luminosity distance:
\begin{eqnarray}
  \log_{10}F_{\rm X}&=& \nonumber \Phi\left(F_{\rm UV},\;D^{\rm cal}_{L}(\Omega_{K},z)\right)\\
  &=&\kappa'+\gamma \log_{10}F_{\rm UV}+2\left(\gamma-1\right)\log_{10}D^{\rm cal}_{L}\;,
\label{eq:Fx}
\end{eqnarray}
where $F_{\rm X}$ and $F_{\rm UV}$ are the rest-frame X-ray and UV fluxes, respectively, and
$\kappa'$ is a constant that contains the slope $\gamma$ and intercept $\kappa$ in
Equation~(\ref{eq:L-relation}), i.e., $\kappa'=\kappa+(\gamma-1)\log_{10}4\pi$.

\begin{figure}
\vskip-0.1in
\centerline{\includegraphics[keepaspectratio,clip,width=0.5\textwidth]{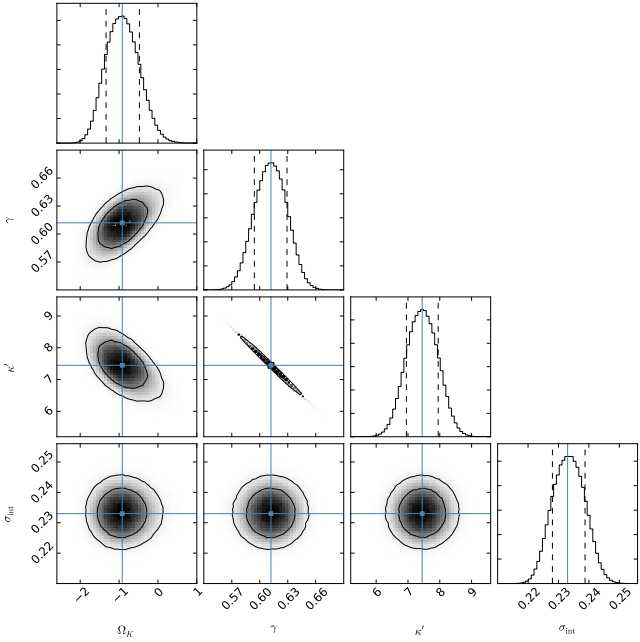}}
\vskip-0.1in
\caption{1-D marginalized probability distributions and 2-D regions with the $1-2\sigma$ contours
corresponding to the cosmic curvature $\Omega_{K}$ and the parameters ($\gamma$, $\kappa'$, and $\sigma_{\rm int}$)
characterizing the luminosity relation. The vertical solid lines represent the best-fits, and the vertical dashed
lines enclose the $1\sigma$ credible region.}
\label{f2}
\end{figure}

One of the limitations we must deal with in using the quasar data, however, is
that the $H(z)$ measurements extend only to $z=2$. As such, we shall employ only a sub-set
of the entire quasar sample that overlaps with the $H(z)$ catalog for the calibration. The
calibration of the luminosity relation will therefore be based only on the 1330 quasars at $z<2$.
The calibrated luminosity relation, along with the curvature parameter $\Omega_{K}$, can be
fitted by maximizing the likelihood function:
\begin{eqnarray}
 \mathcal{L} &=& \nonumber \prod_{i}^{1330}\frac{1}{\sqrt{2\pi}\,\sigma_{{\rm tot}, i}}\;\times \\
            & &\exp\left\{-\,\frac{\left[\log_{10}(F_{\rm X})_{i}-
            \Phi\left(\left[F_{\rm UV}\right]_{i},\;D^{\rm cal}_{L}\left[\Omega_{K},z_{i}
            \right]\right)\right]^{2}}{2\sigma^{2}_{{\rm tot}, i}}\right\}\;,\;\;
\label{eq:likelihood}
\end{eqnarray}
where $\Phi\left(\left[F_{\rm UV}\right]_{i},\;D^{\rm cal}_{L}\left[\Omega_{K},z_{i}\right]\right)$ is
given by Equation~(\ref{eq:Fx}) and the variance
\begin{equation}
\sigma_{\rm tot}^{2}=\sigma_{\rm int}^{2}+\sigma_{\log_{10}F_{\rm X},i}^{2}
+\left[2\left(\gamma-1\right)\frac{\sigma_{D^{\rm cal}_{L},i}}{\ln10\;D^{\rm cal}_{L}
\left(\Omega_{K},z_{i}\right)}\right]^{2}
\label{eq:variance}
\end{equation}
is given in terms of the global intrinsic dispersion $\sigma_{\rm int}$, the measurement uncertainty
$\sigma_{\log_{10}F_{\rm X},i}$ in $\log_{10}(F_{\rm X})_{i}$, and the propagated uncertainty of
$D^{\rm cal}_{L}(\Omega_{K},z_{i})$ derived from cosmic-chronometer measurements. The uncertainty in
$\log_{10}(F_{\rm UV})_{i}$ is presumed to be insignificant compared to the three terms in Equation~(\ref{eq:variance}),
and is therefore ignored in our calculations. In this case, the free parameters are $\sigma_{\rm int}$,
$\gamma$, $\kappa'$, and the curvature parameter $\Omega_{K}$. We use the Python Markove Chain Monte
Carlo (MCMC) module, EMCEE \citep{2013PASP..125..306F}, to get the best-fitting values and their
corresponding uncertainties for these parameters by generating sample points of the probability distribution.
The 1-D marginalized probability distribution for each free parameter and 2-D plots of the $1-2\sigma$
confidence regions for two-parameter combinations are displayed in Figure~\ref{f2}. These contours
show that at the $1\sigma$ level, the optimized parameter values are $\Omega_{K}=-0.918\pm0.429$,
$\gamma=0.612\pm0.017$, $\kappa'=7.447\pm0.500$, and $\sigma_{\rm int}=0.233\pm0.005$. We find that
the measured $\Omega_{K}$ deviates slightly from zero spatial curvature, implying that the current
quasar data favor a mildly closed Universe with a $2.1\sigma$ degree of confidence.

Given the potential impact of such a result, we next consider whether the reconstruction
scheme affects the curvature measurement. We have also performed a parallel comparative analysis of
the discrete cosmic-chronometer data by using a different approach, based on the so-called
Pad$\rm \acute{e}$ approximation.

The Pad$\rm \acute{e}$ approximation to the $H(z)$ function is described by
the rational polynomial (of a specified order)
\begin{equation}
H_{mn}(z)\equiv\frac{{\sum^{m}_{i=0}} a_{i}z^{i}}{1+\sum_{i=1}^{n} b_{i}z^{i}}\;,
\label{eq:Pade}
\end{equation}
where the two non-negative integers ($m$ and $n$) represent the degrees of the numerator and the
denominator, respectively. The coefficients $a_{i}$ and $b_{i}$ can be determined by fitting $H_{mn}(z)$
to the discrete $H(z)$ data. For $z=0$, it is easy to identify $a_{0}\equiv H_0$. There are three
free parameters in the B$\rm \acute{e}$zier polynomial reconstruction. To keep the number of free
parameters the same, we have three different Pad$\rm \acute{e}$ approximations of order
$(m=2,n=0)$, $(m=0,n=2)$, and $(m=1,n=1)$. That is, $H_{20}(z)=a_0+a_1 z+a_2 z^2$,
$H_{02}(z)=a_0[1+b_1z+b_2z^2]^{-1}$, and $H_{11}(z)=[a_0+a_1z][1+b_1z]^{-1}$.

The Pad$\rm \acute{e}$ approximations with $n=0$ actually reduce to Taylor polynomials. We find
that $H_{20}(z)$ fits the data with a reduced $\chi^{2}_{\rm dof}=14.74/28=0.53$, while $H_{02}(z)$
and $H_{11}(z)$ fit the data with an unsatisfactory $\chi^{2}_{\rm dof}=360.84/28=12.89$ and
$\chi^{2}_{\rm dof}=122.87/28=4.39$, respectively. Therefore, we shall consider only the
Pad$\rm \acute{e}$ approximation of order $(2,0)$ for our comparison. The best-fitting
parameters of the reconstructed $H_{20}(z)$ function are $a_{0}=67.76\pm3.69$, $a_1=36.20\pm14.57$,
and $a_2=18.01\pm9.53$ (all in units of km $\rm s^{-1}$ $\rm Mpc^{-1}$). The subsequent steps
to calibrate the distance and measure the curvature are then the same as described above.

Using the Pad$\rm \acute{e}$ based reconstruction, the constraints on the cosmic curvature and
the parameters of the luminosity relation are $\Omega_{K}=-0.930\pm0.430$, $\gamma=0.612\pm0.017$,
$\kappa'=7.452\pm0.501$, and $\sigma_{\rm int}=0.233\pm0.005$. Comparing this inferred $\Omega_{K}$
with that obtained using the B$\rm \acute{e}$zier polynomial reconstruction ($\Omega_{K}=-0.918\pm0.429$),
we see that the adoption of a different reconstruction scheme has only a minimal influence on the
results. For the rest of this paper, we shall therefore adopt the calibrated results based on the
B$\rm \acute{e}$zier polynomial reconstruction.

\begin{figure}
\vskip-0.1in
\centerline{\includegraphics[keepaspectratio,clip,width=0.5\textwidth]{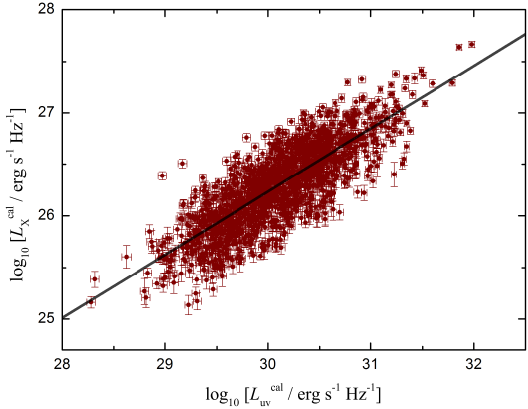}}
\vskip-0.1in
\caption{Calibrated distribution in the $L^{\rm cal}_{\rm UV}$--$L^{\rm cal}_{\rm X}$ plane
for the 1330 quasars at $z<2$. The solid line shows the best-fit result.}
\label{f3}
\end{figure}

The distribution of logarithmic X-ray luminosities, $\log_{10}L^{\rm cal}_{\rm X}$, versus
the UV luminosities, $\log_{10}L^{\rm cal}_{\rm UV}$, is shown in Figure~\ref{f3} for the 1330 quasars
at $z<2$, together with the best-fitting line.
The propagated uncertainties of $\log_{10}L^{\rm cal}_{\rm X}$ and $\log_{10}L^{\rm cal}_{\rm UV}$
are calculated from
\begin{equation}
\sigma_{\log_{10}L^{\rm cal}_{\rm X}}=\left[\sigma^{2}_{\log_{10}F_{\rm X}}+\left(2\frac{\sigma_{D^{\rm cal}_{L}}}{\ln10 D^{\rm cal}_{L}}\right)^{2}\right]^{1/2}
\end{equation}
and
\begin{equation}
\sigma_{\log_{10}L^{\rm cal}_{\rm UV}}=2\frac{\sigma_{D^{\rm cal}_{L}}}{\ln10 \, D^{\rm cal}_{L}}\;,
\end{equation}
respectively.

\begin{figure*}
\vskip-0.2in
\centerline{\includegraphics[keepaspectratio,clip,width=0.9\textwidth]{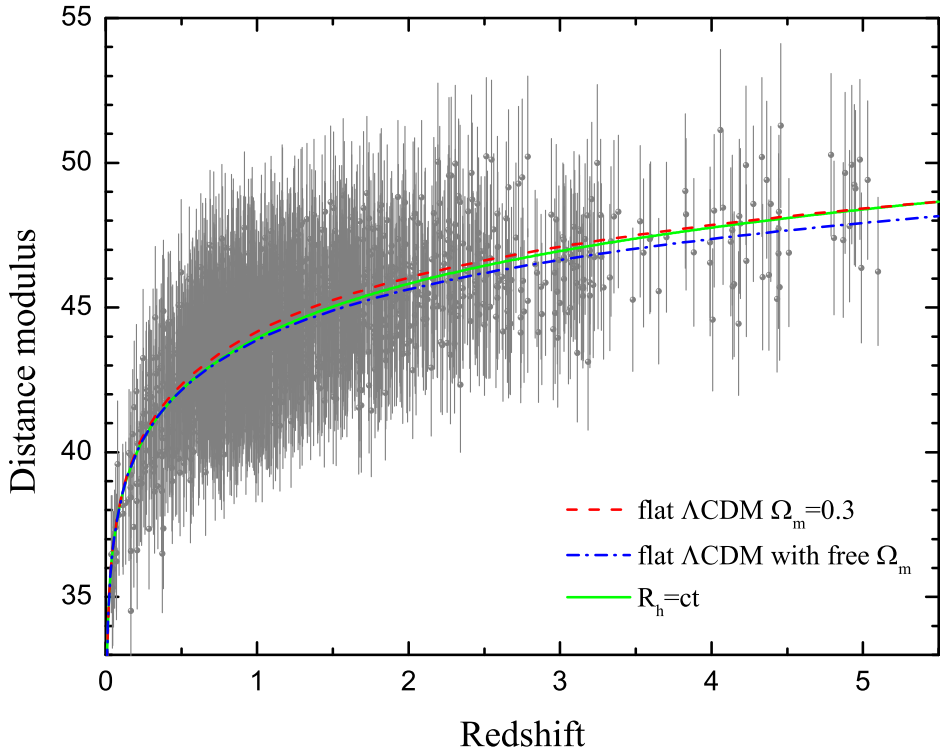}}
\vskip-0.2in
\caption{Hubble diagram for the 1598 calibrated quasars (points), with the $1\sigma$ uncertainties.
The theoretical curves show the concordance flat $\Lambda$CDM model with fixed $\Omega_{\rm m}=0.3$ (dashed),
the optimized flat $\Lambda$CDM model with $\Omega_{\rm m}=0.582^{+0.074}_{-0.059}$ (dot-dashed),
and the $R_{\rm h}=ct$ universe (solid).}
\label{f4}
\end{figure*}

\cite{2015ApJ...815...33R} found no evidence of a redshift evolution for the luminosity relation.
We shall therefore assume that the optimized relation we have derived at $z< 2$ holds at all
redshifts with the same slope and intercept. Extrapolating the calibrated luminosity relation to
high-$z$ ($z>2$), we then derive the distance moduli of the whole quasar sample, including 1598 sources.
With the calibrated luminosity relation, the distance modulus of a quasar can be obtained using
\begin{equation}
\mu_{\rm obs}=\frac{5}{2\left(\gamma-1\right)}\left[\log_{10}F_{\rm X}-\gamma\log_{10}F_{\rm UV}-\kappa'\right]-97.447\;.
\end{equation}
The error $\sigma_{\mu_{\rm obs}}$ in $\mu_{\rm obs}$ is calculated via error propagation, i.e.,
\begin{equation}
\sigma_{\mu_{\rm obs}}=\mu_{\rm obs}\left[\left(\frac{\sigma_{\gamma}}{\gamma-1}\right)^{2}
+\left(\frac{\sigma_{y}}{\log_{10}F_{\rm X}-\gamma\log_{10}F_{\rm UV}-\kappa'}\right)^{2}\right]^{1/2},
\end{equation}
where $\sigma_{y}=\left[\sigma^{2}_{\log_{10}F_{\rm X}}+\left(\sigma_{\gamma}\log_{10}F_{\rm UV}\right)^{2}+
\sigma^{2}_{\kappa'}+\sigma^{2}_{\rm int}\right]^{1/2}$, and $\sigma_{\gamma}$ and $\sigma_{\kappa'}$ are
the uncertainties of the slope $\gamma$ and intercept $\kappa'$. The calibrated quasar Hubble diagram is
shown in Figure~\ref{f4}.

\section{Testing Cosmological Models}
\label{sec:constraint}
In this section, we use the calibrated quasar distance moduli to test certain cosmological models.
The cosmological parameters are optimized by minimizing the $\chi^{2}$ statistic, i.e.,
\begin{equation}
\chi^{2}=\sum_{i}^{1598}\frac{\left[\mu_{{\rm obs},i}-\mu_{\rm th}\left(z_{i}\right)\right]^{2}}{\sigma^{2}_{\mu_{\rm obs},i}}\;,
\end{equation}
where $\mu_{\rm th}\equiv5\log_{10}\left[D_{L}(z)/{\rm Mpc}\right]+25$ is the theoretical distance modulus
of a quasar at redshift $z$. The determination of $D_{L}$ requires the assumption of a particular cosmological model.
For the sake of consistency, we adopt the Hubble constant as the best-fitting value derived from the model-independent
analysis of the cosmic-chronometer data ($H_0=67.76$ km $\rm s^{-1}$ $\rm Mpc^{-1}$) in the optimization procedure.
Here we discuss how the fits have been optimized for $\Lambda$CDM and $R_{\rm h}=ct$. The outcome for each model
is carefully described and discussed in subsequent sections.

\begin{figure}
\vskip-0.1in
\centerline{\includegraphics[keepaspectratio,clip,width=0.6\textwidth]{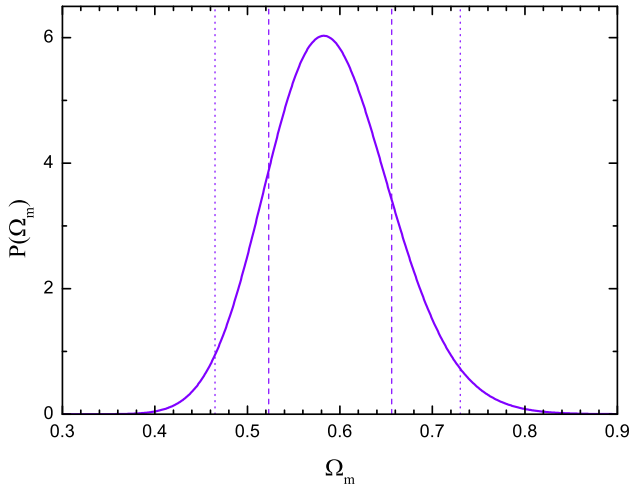}}
\vskip-0.1in
\caption{Probability distribution function of the matter density parameter $\Omega_{\rm m}$.
The $1\sigma$ and $2\sigma$ credible intervals are indicated by the vertical dashed and dotted lines, respectively.}
\label{f5}
\end{figure}

\subsection{$\Lambda$CDM}
In a flat $\Lambda$CDM universe with zero spatial curvature, the theoretical luminosity distance is given as
\begin{equation}
D^{\rm \Lambda CDM}_{L}\left(z\right)=\frac{c}{H_0}\left(1+z\right)\int_{0}^{z}\frac{dz'}{\sqrt{\Omega_{\rm m}\left(1+z'\right)^{3}+\Omega_{\Lambda}}}\;,
\end{equation}
where $\Omega_{\rm m}$ is the scaled matter density today and $\Omega_{\Lambda}=1-\Omega_{\rm m}$
is the cosmological constant energy density. (We ignore the contribution from radiation, which is
insignificant compared to that of matter and dark energy in this redshfit range.) The Hubble constant
$H_0$ is fixed to be the value obtained from the model-independent analysis of the discrete $H(z)$
data, so the sole remaining parameter in flat $\Lambda$CDM is $\Omega_{\rm m}$. With the calibrated
distance moduli of quasars, the resulting constraint on $\Omega_{\rm m}$ is shown in Figure~\ref{f5}.
The best-fitting value is $\Omega_{\rm m}=0.582^{+0.074}_{-0.059}$ at the $1\sigma$ confidence level.
With $1598-1=1597$ degrees of freedom, the reduced $\chi^{2}$ is $\chi^{2}_{\rm dof}=648.57/1597=0.41$.

This optimized $\Omega_{\rm m}$ value is, however, in some tension with that inferred from other kinds
of data. In particular, the concordance $\Lambda$CDM model with $\Omega_{\rm m}\approx 0.3$ seems to account
best for many other cosmological observations \citep{2015PhRvD..92l3516A,2018arXiv180706209P,2018ApJ...859..101S}.
It must be emphasized, however, the $\Lambda$CDM is rarely tested in the redshift range between the farthest observed
SNe Ia and the last scattering surface (producing the CMB). Recently, \cite{2019NatAs...3..272R} showed that
a $\sim4\sigma$ tension exists between the high-$z$ data and the $\Lambda$CDM model, based on a model-independent
parametrization of the Hubble diagram using these sources. Fitting the calibrated distance moduli of 1598 quasars
obtained from our model-independent technique with the concordance model using $\Omega_{\rm m}=0.3$, we obtain a
$\chi^{2}$ per degree of freedom of $\chi^{2}_{\rm dof}=676.60/1598=0.42$.

These results are somewhat consistent with those of \cite{2019NatAs...3..272R}, in the sense
that the concordance model does not provide the best fit to these quasar data extending up to $z\sim 6$,
when compared to other formulations and/or parameter values. Having said this, the errors reported for the
data appear to be larger than one would expect from their scatter, which is probably why our inferred $\chi^2_{\rm dof}$'s
are much smaller than $1$. Simply on the basis of the reduced $\chi^2$, the concordance model fits the data
quite well. When compared to the other two fits reported in Table~\ref{table1}, however, one can see, especially from Figure~\ref{f5},
that the concordance model with $\Omega_{\rm m}=0.3$ is disfavored at the $4.3\sigma$ level, somewhat confirming
the result of \cite{2019NatAs...3..272R}.

\subsection{The $R_{\rm h}=ct$ universe}
The luminosity distance in the $R_{\rm h}=ct$ universe
\citep{2003eisb.book.....M,2007MNRAS.382.1917M,2013A&A...553A..76M,2012MNRAS.419.2579M,Wei2015},
is given as
\begin{equation}
D^{R_{\rm h}=ct}_{L}\left(z\right)=\frac{c}{H_0}\left(1+z\right)\ln\left(1+z\right)\;.
\end{equation}
The $R_{\rm h}=ct$ cosmology has only one free parameter, $H_0$, but since the Hubble constant is fixed,
there are no free parameters left to fit the quasar Hubble diagram. The $\chi^{2}$ per degree of
freedom of this cosmology is $\chi^{2}_{\rm dof}=655.33/1598=0.41$.

To facilitate a direct comparison between $R_{\rm h}=ct$ and $\Lambda$CDM, we show in Figure~\ref{f4}
the best-fitting theoretical curves for the concordance flat $\Lambda$CDM model with fixed $\Omega_{\rm m}=0.3$ (dashed),
the optimized flat $\Lambda$CDM model with $\Omega_{\rm m}=0.582^{+0.074}_{-0.059}$ (dot-dashed), and the $R_{\rm h}=ct$
universe (solid). On the basis of their $\chi^{2}_{\rm dof}$ values, the optimized $\Lambda$CDM model and the $R_{\rm h}=ct$
universe appear to fit the data comparably well.
Because these models have different numbers of free parameters, however, a simple $\chi^{2}_{\rm dof}$-minimization is not
sufficient to fairly judge which is a better match to the data. A comparison of the likelihoods indicating which
cosmology is closest to the `true' model must be based on model selection criteria, which we discuss next.

\subsection{Statistical Performance with Quasars}
Since the sample we have is very large, we apply the most appropriate model selection tool---the Bayes Information Criterion
(BIC; \citealt{1978AnSta...6..461S}) to test the statistical performance of the models:
\begin{equation}
{\rm BIC}=\chi^{2}+\left(\ln N\right)f\;,
\end{equation}
where $N$ is the number of data points and $f$ is the number of free parameters.  The BIC is a large-sample
($N\rightarrow\infty$) approximation to the outcome of a conventional Bayesian inference procedure for
deciding between models. Among the models being tested, the one with the least BIC score is the one most
preferred by this criterion. A more quantitative ranking of models can be computed as follows. With
${\rm BIC}_\alpha$ characterizing model $\alpha$, the unnormalized confidence that this model is correct
is the `Bayes weight' $\exp(-{\rm BIC_\alpha}/2)$. The relative likelihood of model $\alpha$ being correct is then
\begin{equation}
P(\alpha)=\frac{\exp(-{\rm BIC_\alpha}/2)}{\sum_{i}\exp(-{{\rm BIC}_{i}}/2)}\;,
\label{eq:BIC}
\end{equation}
where the sum in the denominator runs over all of the models being tested simultaneously. The outcome of this
analysis is summarized in Table~1. According to these results, we conclude that $R_{\rm h}=ct$ is slightly
preferred over $\Lambda$CDM with a likelihood of $\sim57.69\%$ versus 42.31\%. The concordance model with a fixed
$\Omega_{\rm m}=0.3$ can be safely discarded as having a probability of only $\sim10^{-5}$ of being correct compared
to the other two fits reported here. To facilitate the comparison, Table~\ref{table1} also shows the individual BIC
values and each model's relative likelihood.

\section{Summary and Discussion}
\label{sec:summary}
In this work, we have proposed a method for calibrating the luminosity distance in a model-independent
way, and using this to measure the spatial curvature. This approach is achieved by combining observations
of quasars and cosmic chronometers. First, we use the discrete cosmic-chronometer measurements to
reconstruct the continuous Hubble function $H(z)$ using a polynomial fit. The co-moving distance function
can then be derived by directly calculating the integral of the reconstructed $H(z)$ function. With the
curvature parameter $\Omega_{K}$ taken into account, we can transform the co-moving
distance into the curvature-dependent luminosity distance $D^{\rm cal}_{L}(\Omega_{K},z)$.
Finally, based on the X-ray versus UV luminosity correlation for quasars in the redshift
range overlapping with $H(z)$, we combine the redshift and flux measurements of 1330 sources
at $z<2$ with $D^{\rm cal}_{L}(\Omega_{K},z)$ to constrain
both the curvature parameter and the parameters characterizing the luminosity relation in
a model-independent way.

\begin{table}
\centering \caption{Best-fitting results in different cosmological models with the whole sample of 1598 quasars}
\begin{tabular}{lcccc}
\hline
\hline
 Model &  $\Omega_{\rm m}$  &  $\chi^{2}_{\rm dof}$  &  BIC  & Likelihood \\
\hline
Concordance     &   0.3 (fixed)       &   0.42   &   676.60  &  1E-5   \\
$\Lambda$CDM    &   $0.582^{+0.074}_{-0.059}$   &   0.41   &   655.95  & 42.31\%   \\
$R_{\rm h}=ct$  &   --                &   0.41   &   655.33  & 57.69\%   \\
\hline
\end{tabular}
\label{table1}
\end{table}

\begin{table}
\centering \caption{Best-fitting results in different cosmological models using only the 1330 quasars at $z<2$}
\begin{tabular}{lcccc}
\hline
\hline
 Model &  $\Omega_{\rm m}$  &  $\chi^{2}_{\rm dof}$  &  BIC  & Likelihood \\
\hline
Concordance     &   0.3 (fixed)       &   0.43   &   568.62  &  0.02\%    \\
$\Lambda$CDM    &   $0.580^{+0.087}_{-0.068}$   &   0.41   &   556.18  & 10.15\%   \\
$R_{\rm h}=ct$  &   --                &   0.41   &   551.82 & 89.83\%   \\
\hline
\end{tabular}
\label{table2}
\end{table}

This analysis suggests that the curvature parameter is constrained to be
$\Omega_{K}=-0.918\pm0.429$, which deviates slightly from zero. That is, the current quasar
data appear to favor a mildly closed Universe at the $2.1\sigma$ level. The optimized
correlation parameters are $\gamma=0.612\pm0.017$, $\kappa'=7.447\pm0.500$, and
$\sigma_{\rm int}=0.233\pm0.005$. Assuming a standard flat $\Lambda$CDM model with
$\Omega_{\rm m}=0.3$, \cite{2019NatAs...3..272R} found the optimized values of the
correlation parameters to be $\gamma=0.633\pm0.002$ and $\sigma_{\rm int}=0.24$. Our
constraints are very similar to those of \cite{2019NatAs...3..272R}, though not exactly
the same. This comparison between the two approaches attests to the reliability of our
calculation, but also indicates the importance of developing a cosmology-free calibration.

Assuming that the extrapolation of the calibrated luminosity relation beyond $z\sim2$ is valid,
we obtained a new sample of distance moduli for the 1598 different quasars, and used them
to compare two competing cosmological scenarios, i.e., $\Lambda$CDM and the $R_{\rm h}=ct$ universe.
We showed that the latter fits the data with a reduced $\chi^{2}_{\rm dof}=655.33/1598=0.41$.
By comparison, the optimal flat $\Lambda$CDM model fits these same data with a reduced
$\chi^{2}_{\rm dof}=648.57/1597=0.41$ for a matter density parameter $\Omega_{\rm m}=0.582^{+0.074}_{-0.059}$.
The model comparison shows that $R_{\rm h}=ct$ is slightly preferred over $\Lambda$CDM with
a likelihood of $\sim57.7\%$ versus 42.3\% when $\Omega_{\rm m}$ is allowed to deviate from
its concordance value. $R_{\rm h}=ct$ is much more strongly preferred over $\Lambda$CDM,
however, when the latter is based on the concordance parameter values.

To examine whether the calibrated luminosity relation can reliably be extrapolated to high-$z$,
we also used just the restricted sample of 1330 calibrated quasar distance moduli at $z<2$ to compare
different cosmological models. The results are shown in Table~\ref{table2}. The likelihoods indeed change
somewhat, but not qualitatively. $R_{\rm h}=ct$ is more strongly favoured over $\Lambda$CDM with a
likelihood of $\sim89.8\%$ versus 10.2\%. The outcomes based on the reduced and complete samples
are therefore consistent with each other.

\acknowledgments
We are grateful to the anonymous referee for useful comments that have
helped us improve the presentation of the paper.
This work is partially supported by the National Natural Science Foundation of China
(grant Nos. 11603076, 11673068, 11725314, and U1831122), the Youth Innovation Promotion
Association (2017366), the Key Research Program of Frontier Sciences (grant Nos. QYZDB-SSW-SYS005
and ZDBS-LY-7014), the Strategic Priority Research Program ``Multi-waveband gravitational wave universe''
(grant No. XDB23000000) of Chinese Academy of Sciences, the ``333 Project'' of Jiangsu Province,
and the Guangxi Key Laboratory for Relativistic Astrophysics.


\end{document}